\documentclass[preprint,superscriptaddress,amsmath,amssymb,aps,prl]{revtex4-1}
\usepackage{amsmath}
\usepackage{amssymb}
\usepackage{graphicx}
\usepackage{braket}
\usepackage{bm}
\usepackage{tikz}

\begin{document}

	\title{Nonlinear magnetoelectric effect in atomic vapor}
	
	\author{Sushree S. Sahoo}
	\email{ssahoo@tifrh.res.in}
	\affiliation{National Institute of Science Education and Research Bhubaneswar, HBNI, Jatni-752050, India.}
	\affiliation{TIFR Centre for Interdisciplinary Sciences, Tata Institute of Fundamental Research, Hyderabad-500017, India.}
	\author{Soumya R. Mishra}
	\affiliation{National Institute of Science Education and Research Bhubaneswar, HBNI, Jatni-752050, India.}
	\author{G. Rajalakshmi}
	\affiliation{TIFR Centre for Interdisciplinary Sciences, Tata Institute of Fundamental Research, Hyderabad-500017, India.}
	\author{Ashok K. Mohapatra}
	\email{a.mohapatra@niser.ac.in}
	\affiliation{National Institute of Science Education and Research Bhubaneswar, HBNI, Jatni-752050, India.}

\begin{abstract}
{\bf Magnetoelectric (ME) effect refers to the coupling between electric and magnetic fields in a medium resulting in electric polarization induced by magnetic fields and magnetization induced by electric fields~\cite{Dell70,Fieb05}. The linear ME effect in certain magnetoelectric materials such as multiferroics has been of great interest due to its application in the fabrication of spintronics devices, memories, and magnetic sensors~\cite{Toku07,Klee13,Scot07,Bibe08,Spal19}. However, the exclusive studies on the nonlinear ME effect are mostly centered on the investigation of second-harmonic generation in chiral materials~\cite{Maki95, Bote05, Fie05}. Here, we report the demonstration of nonlinear wave mixing of optical electric fields and radio-frequency (rf) magnetic fields in thermal atomic vapor, which is the consequence of the higher-order nonlinear ME effect in the medium. The experimental results are explained by comparing with density matrix calculations of the system. We also experimentally verify the expected dependence of the generated field amplitudes on the rf field magnitude as evidence of the magnetoelectric effect. This study can open up the possibility for precision rf-magnetometry due to its advantage in terms of larger dynamic range and arbitrary frequency resolution.}

\end{abstract}


\maketitle

The electrical polarization due to magnetoelectric (ME) effect induced in a medium in response to the applied electric field $E$ and magnetic field $B$ is defined by the general expression, $P_i({ E},{ B})=\chi^{ee}_{ij}E_j+\chi^{em}_{ij} B_{j}+\chi^{emm}_{ijk} B_j B_k+\chi^{eem}_{ijk}E_jB_k+\chi^{eemm}_{ijkl}E_jB_kB_l+...$, where the indices $ijk$ refer to the polarisation components of the fields whereas the indices $e$ and $m$ denote the electric and magnetic fields respectively. $\chi^{ee}_{ij}$ signifies the linear electric susceptibility, $\chi^{em}_{ij}$ describes the linear ME effect while the leading higher-order ME contributions are described by the tensors $\chi^{emm}_{ijk}$, $\chi^{eem}_{ijk}$ and $\chi^{eemm}_{ijkl}$. In this study, we explore the nonlinear polarization terms given by, $P^{(2)}_i=\chi^{eem}_{ijk}E_jB_k$ and $P^{(3)}_i=\chi^{eemm}_{ijkl}E_jB_kB_l$. The polarisation, $P^{(2)}_i$ is a result of mixing of three fields i.e. two input fields (one electric and one magnetic field) and one generated electric field whereas $P^{(3)}_i$ results from mixing of four fields i.e. three input fields (one electric and two magnetic fields) and one generated electric field. The mixing between microwave and optical fields in atomic systems is an example of such mixing processes~\cite{Zibr02,Adwa19}.

In this work, we demonstrate the nonlinear ME effects achieved through the parametric interaction of optical and rf fields via multi-wave mixing processes resulting in the efficient generation of optical fields. The studies on the interaction of optical and rf fields so far are based on the induced spin polarization in a system by an rf field while coupling the Zeeman sublevels. This leads to a polarisation rotation of an input linearly polarised light traversing the medium~\cite{Savu05,Lee06,Ledb07,Zigd10,Chal12,Kede14,Cohe19}. We couple one of the ground states of the atomic system to an excited state using an optical field while the rf field couples the neighboring Zeeman sublevels of the ground state such that it induces ground-state coherence in the system facilitating the mixing process. This results in the system producing light at optical frequencies as satisfied by the energy conservation due to the process. We also study the characteristic features of the generated fields such as polarization, resonance width, and the variation of generation amplitudes with input optical power. 

\begin{figure}[t]
	\centering
	\includegraphics[width=100mm,scale=1]{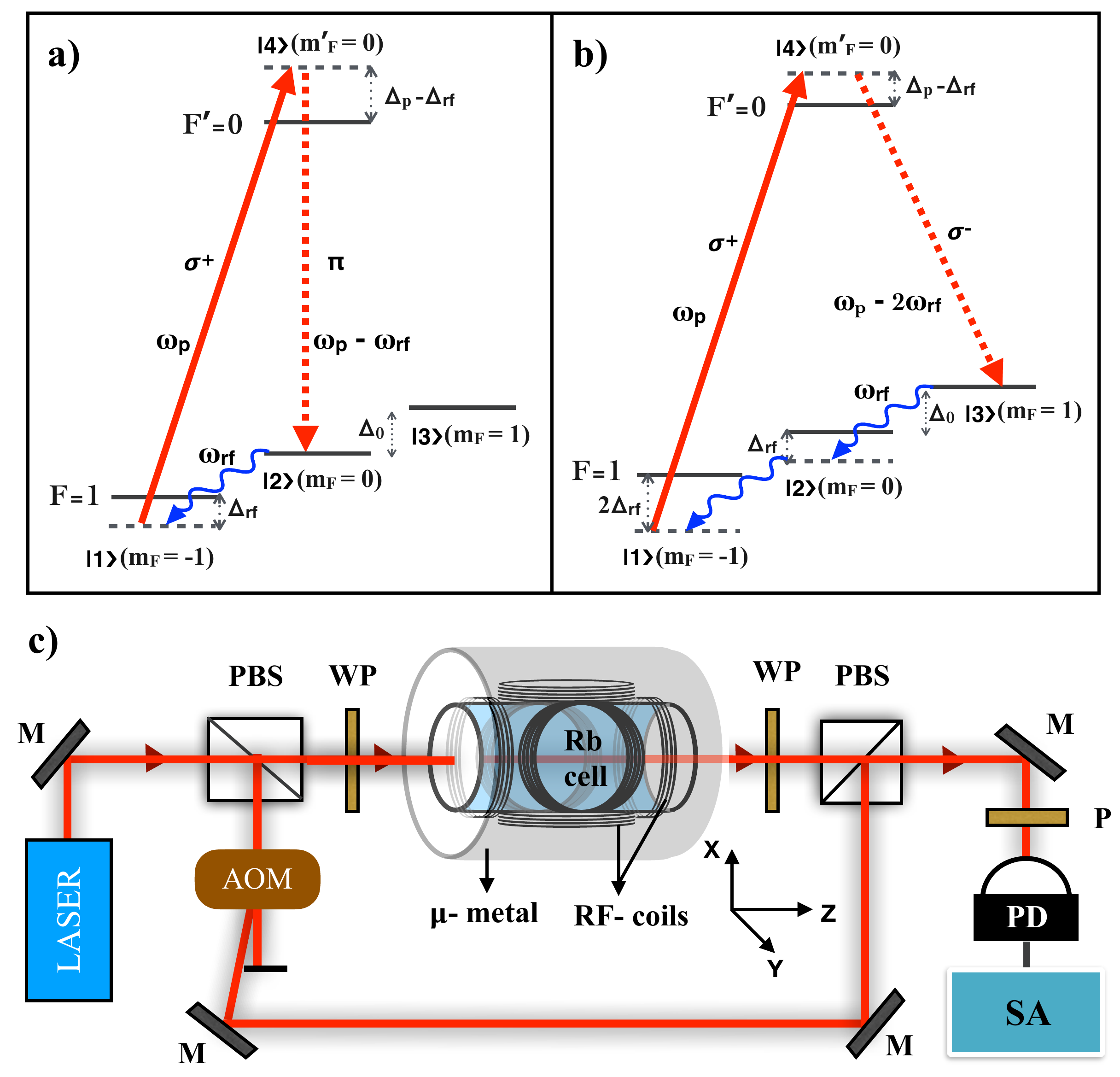}
	\caption{Depiction of the (a) three-wave and (b) four-wave mixing processes in the schematic energy diagram for $D_2$ line, $^{87}$Rb $F = 1\rightarrow F=0$ transition. Here, the input fields are the pump ($\omega_p$) and the rf field ($\omega_{rf}$) leading to the optical generation at frequencies, $\omega_p-\omega_{rf}$ via three-wave mixing and $\omega_p-2\omega_{rf}$ via four-wave mixing processes. $\Delta$ ($\Delta_{rf}$) is the detuning of the input optical (rf) field from the corresponding atomic transition. c) Schematic of the experimental setup for the observation of the mixing process. PBS: Polarising beam splitter, M: Mirror, AOM: Acousto-Optic Modulator, WP: Wave plate, P: Polariser, PD: Photo-detector, SA: Spectrum analyzer}
	\label{b1}	
\end{figure}

The schematic of the atomic energy levels coupled by the input optical field and rf magnetic field is shown in Fig.~\ref{b1} (a) and (b). The pump field ($\omega_p$) of $\sigma^{+}$ polarisation, coupling the ground state with $m_F=-1$ to the excited state with $m^{'}_F=0$ drives the population from  $m_F=-1$ to $m_F=0$ and $m_F=1$ ground states via optical pumping~\cite{Happ72}. There are two possible parametric cycles in the system. An atom present in $m_F=0$ ground state emits one $\sigma^+$ rf photon to come to $m_F=-1$ state, then absorbs the $\sigma^+$ pump photon to be excited to $m^{'}_F=0$ state and finally emits a $\pi$ optical photon to come back to $m_F=0$ state. This parametric process is a three-wave mixing process, which can be described by $P_{\pi}^{(2)}(=\chi_{\pi\sigma^{+}\sigma^{+}}^{eem}E_{\sigma^{+}}B^*_{\sigma^{+}})$ as discussed before. Similarly, in the four-wave mixing process, the atom starting with $m_F=1$ ground state emits two $\sigma^+$ rf photons to come to $m_F=-1$ state and absorbs one $\sigma^+$ pump photon to be excited to $m^{'}_F=0$ state and then comes back to $m_F=1$ state by emitting a $\sigma^-$ optical photon. This four-wave mixing process is described by  $P_{\sigma^{-}}^{(3)}(=\chi_{\sigma^{-}\sigma^{+}\sigma^{+}\sigma^{+}}^{eemm}E_{\sigma^{+}}B_{\sigma^{+}}^{*2})$.  The energy conservation leads to optical field generation at frequencies, $\omega_{g1} (=\omega-\omega_{rf}$) and $\omega_{g2}(=\omega-2\omega_{rf}$) via three-wave mixing and four-wave mixing processes respectively. The polarization states of the generated fields are decided by the angular momentum conservation in both the processes i.e. for an input $\sigma^{+}$ polarised pump beam, the three-wave (four-wave) mixing process leads to generated beam with $\pi$ ($\sigma^{-}$) polarisation. Furthermore, as the wave vector of the rf field is negligible compared to that of the optical field, the phase-matching conditions ensure that the direction of the generated beams is same as that of the input pump beam.

\begin{figure}[t]
	\begin{center}
		\includegraphics[width=145mm,scale=1]{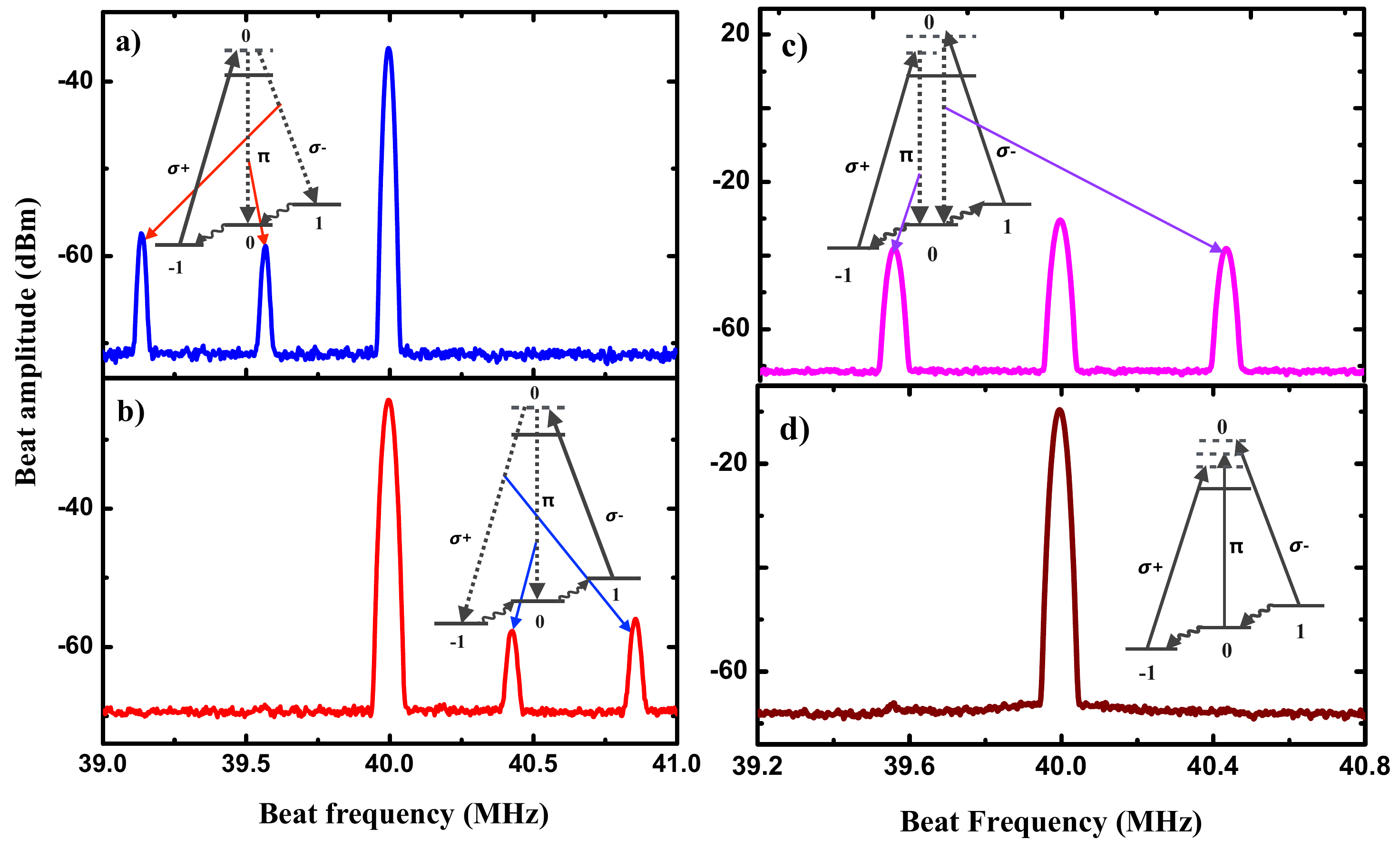}
		\caption{Experimental data of the generated beat amplitudes when the input pump beam is a) $\sigma^+$ polarized, b) $\sigma^-$ polarized, c) linearly polarized with equal components of $\sigma^+$ and $\sigma^-$ polarization ($\frac{1}{\sqrt{2}}(|\sigma^+\rangle+|\sigma^-\rangle )$ and d) linearly polarized with equal components of all $\sigma^+$, $\sigma^-$ and $\pi$ polarizations. The larger peak at 40 MHz refers to the beat note corresponding to the interference of the LO and the input pump light. The other peaks corresponding to the generated light fields are indicated by the inset showing the respective wave mixing processes in the energy-level diagrams.}
		\label{b2}
	\end{center}
\end{figure}

The schematic of the experimental set-up is depicted in Fig~\ref{b1}(c). The pump field propagates through a magnetically shielded rubidium vapor cell and the generated fields are analyzed by heterodyne detection, combining with a local oscillator (LO) (Refer: Methods). We use three pairs of Helmholtz coils to apply the magnetic field along x-, y- and z-directions. In the first experiment, we apply a static magnetic field along the z-direction and rf magnetic field along the y-direction. The experimental data for the input circularly polarised light is presented in Fig.~\ref{b2} (a) and (b). As expected, when we use the pump with $\sigma^+$ ($\sigma^-$) circular polarization, the optical fields due to the mixing process are generated with lower (higher) optical frequencies than the pump field and hence the interference peak with LO appears at the left (right) side of the main peak. It is interesting to note that this frequency up/down conversion process is a direct method to determine the handedness of circular polarization of light interacting with the medium.

\begin{figure*}[t]
	\begin{center}
		\includegraphics[width=165mm,scale=1]{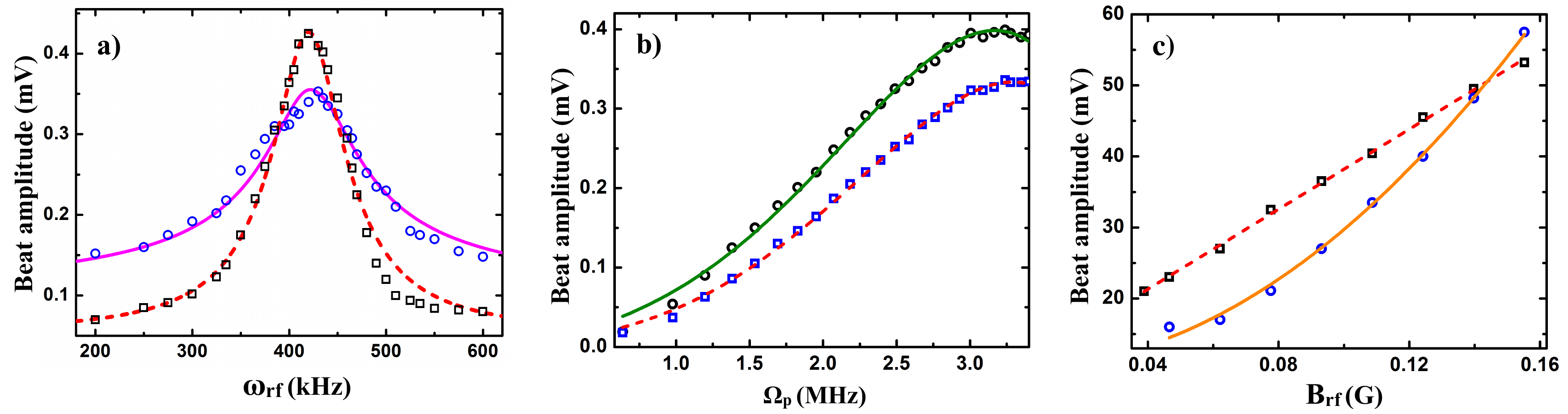}
		\caption{a) Resonance curves corresponding to the generated beat signals as a variation of the input rf frequency ($\omega_{rf}$). Here, blue open circles (black open squares) depict the experimental data, and magenta solid line (dashed red line) depicts the theoretical fitting corresponding to the generated field $\omega_{g1}$ ($\omega_{g2}$). The input parameters for the model are, $\Omega_p=0.5$ MHz, $\Omega_{rf}=80$ kHz, $\Delta_p=150$ MHz, $\Gamma=6$ MHz and $\Delta_{0}=425$ kHz whereas the fitting parameter is $\gamma=65$ kHz. b) Variation of the generated beat amplitudes at resonance as a variation of the input pump Rabi frequency ($\Omega_p$). The blue open circles (black open squares) depict the experimental data and the magenta solid line (dashed red line) shows the theoretical fitting corresponding to $\omega_{g1}$ ($\omega_{g2}$). Here, the fitting function is of the form, $a\Omega_p+b\Omega_p^3+c\Omega_p^5$ with the fitting parameters given as, $a=31.72\pm1.88$ $(52.34\pm 3.54)$, $b=17.53\pm0.56$ $ (20.86\pm1.05)$ and $c=-1.02\pm0.04$ $(-1.35\pm 0.07)$ for $\omega_{g1}$ ($\omega_{g2}$) signal. c) Variation of beat amplitudes of the generated fields with the amplitude of rf magnetic field ($B_{rf}$) showing linear (black open squares) and quadratic (blue open circles) behavior for $\omega_{g1}$ and $\omega_{g2}$ signals respectively. The experimental data are fitted with the corresponding functional form as shown by the red dashed line (linear) and orange solid line (quadratic).}
		\label{b4}
	\end{center}
\end{figure*}

In the further experiment, the static and rf magnetic fields are applied along y- and z-directions respectively. The linear polarisation of the input pump field along the x-direction is a linear combination of $\sigma^+$ and $\sigma^-$ polarisations i.e. of the form, $\frac{1}{\sqrt{2}}(\left|\sigma^+\right\rangle+\left|\sigma^-\right\rangle)$. Both the generated light fields in this case are $\pi$-polarised with optical frequencies, $\omega_p+\omega_{rf}$ and $\omega_p-\omega_{rf}$.  The respective experimental data for the beat amplitudes are presented in Fig.~\ref{b2}(c). On the other hand, when the input pump is $\pi$-polarised, the generated beams are observed to be $\sigma^+$ ($\omega_p+\omega_{rf}$) and $\sigma^-$ ($\omega_p-\omega_{rf}$) polarized. Furthermore, when the input linear polarization is such that it has equal components of $\sigma^+$, $\sigma^-$, and $\pi$ polarizations, all the ground-states become equally populated leading to no generation due to mixing. In this case, the ground-state coherence, which is responsible for the efficient mixing process is no longer induced in the system as it requires a non-zero population difference between the Zeeman sublevels. It corresponds to the vanishing beat amplitudes for the generated fields as depicted in Fig.~\ref{b2} (d).

The mixing process and hence the light generation is most efficient when the frequency of the rf field ($\omega_{rf}$) matches the Zeeman splitting ($\Delta_0$). In the experiment, the rf magnetic field ($\omega_{rf}$) is scanned around the Zeeman splitting to observe the resonance curves for both $\omega_{g1}$ and $\omega_{g2}$ for the case of the circularly polarised input pump beam. Fig.~\ref{b4}(a) shows the experimental data for the resonance curves peaked at $425$ kHz. We use the expression of $\chi_{(g1)}^{eem}$ and $\chi_{(g2)}^{eemm}$ from the theoretical model (refer: Methods) to fit the experimental data. From the fitting, we find the value of $\gamma$ to be $65$ kHz which is the dephasing rate and is mostly dominated by the transit time of the atoms through the laser beams and the magnetic inhomogeneity present in the medium.

We also study the beat amplitude of the generated beams by varying the input pump Rabi frequency ($\Omega_p$) and the experimental data for both $\omega_{g1}$ and $\omega_{g2}$ are presented in Fig.~\ref{b4}(b). To model this experimental observation, we consider the propagation equation for the generated field with Rabi frequency $\Omega_{gi} (i=1,2)$, which  can be written as, $\frac{d\Omega_{gi}}{dz}= -\alpha_{gi}\Omega_{gi}+\kappa_{gi}$ where $\alpha_{gi}=\frac{k_{gi}}{2}\text{Im}(\chi_{eff}^{(1)})$ corresponds to the gain/absorption in the medium whereas $\kappa_{g1}=i\frac{\hbar}{\mu}k_{g1}\chi^{eem}_{(g1)}$ and  $\kappa_{g2}=i\frac{\hbar}{\mu}k_{g2}\chi^{eemm}_{(g2)}$ (Refer: Methods), where $k_{g1}$ and $k_{g2}$ are magnitude of the wave vectors corresponding to the generated fields, $\Omega_{g1}$ and $\Omega_{g2}$ respectively. This equation can be solved using the initial condition; $\Omega_{gi}=0$ at $z=0$ to get, $\Omega_{gi}=\frac{\kappa_{gi}}{\alpha_{gi}}(1-e^{-\alpha_{gi} l})$ where $l$ is the length of the vapor cell. Using the linear dependence of $\kappa_{gi}$ with $\Omega_p$ and expanding $\alpha_{gi}=\alpha_0+\alpha_1\Omega_p^2+\alpha_2\Omega_p^4$ with $\alpha_0$, $\alpha_1$ and $\alpha_2$ being the absorption/gain co-efficients corresponding to the linear and nonlinear processes, the expression for $\Omega_{gi}$ can be simplified under the assumption of $\alpha_{gi} l \ll 1$ to a polynomial of odd orders of $\Omega_p$ in the form, $\Omega_{gi}= a\Omega_p+b\Omega_p^3+c\Omega_p^5$. We use this functional form to fit the experimental data for Fig.~\ref{b4}(b), which shows a good matching between the model and experiment.

The wave-mixing process is found to be efficient even when the amplitude of the input rf field is very small. In this limit, the expressions for the susceptibilities of the generated fields can be simplified using the approximation, $\Omega_{rf} << \gamma$ as (refer: Methods), 
$\chi^{eem}_{(g1)}= \frac{N\mu_{42}^2}{\epsilon_0\hbar\Omega_{g1}}\frac{\Omega_p \Omega_{rf}}{2\Delta_p (\Delta_{rf}+i\gamma)}$ and
$\chi^{eemm}_{(g2)}= -\frac{N\mu_{43}^2}{\epsilon_0\hbar\Omega_{g2}}\frac{\Omega_p \Omega_{rf}^2}{2\Delta_p (\Delta_{rf}+i\gamma)(2\Delta_{rf}+i\gamma)}$. These analytical expressions clearly depict the linear dependence of $\chi^{eem}_{(g1)}$ as well as the quadratic dependence of $\chi^{eemm}_{(g2)}$ on $\Omega_{rf}$. To verify this dependence, we experimentally measured the generated beat amplitudes as a function of the input rf field amplitude (B$_{rf}$). The corresponding experimental data with linear and quadratic fittings are presented in Fig.~\ref{b4} (c). This experimental observation is a further confirmation of the nonlinear mixing between the optical field and the rf magnetic field occuring in the system.

We have also observed that at low rf frequency i.e. below 100 kHz, another nonlinear process known as the forward four-wave mixing~\cite{Saho17} becomes dominant for input linearly polarized pump beam. In this case, the generated fields due to rf and optical mixing act as a seed for the all-optical forward four-wave mixing process and hence results in large amplification of the signal. These results can have profound applications in the field of rf magnetometry. Previous works on rf magnetometry are based on atomic magnetometers or SQUIDs. RF SQUIDs are less sensitive to their DC counterparts and give sensitivities of the order of 30fT/ Hz$^{1/2}$ at 77K \cite{Brag95}. Atomic magnetometers measure the polarisation rotation of an input linearly polarised beam in presence of an rf field coupling the Zeeman sublevels of the ground state and so far the sensitivity reached is 0.3 fT/Hz$^{1/2}$ at 0.5 MHz in thermal vapor~\cite{Kede14} and 330 pT/ Hz$^{1/2}$ in the cold atomic ensemble~\cite{Cohe19}. In our case, the rf magnetic field sensitivity is anticipated to be of similar order as we exploit the ground-state coherence induced by the rf field, and hence, its effect should be comparable to that due to the ground-state population distribution induced by the rf field~\cite{Chal12, Kede14}. Moreover, the parametric nature of the basic process in our system promises sensitivity with a larger dynamic range.  Another interesting feature of the proposed rf magnetometer using our system is its arbitrary resolution in frequency, which is limited by the resolution of the measurement device employed. Therefore, the system has the potential to surpass the state-of-the-art frequency resolution of millihertz in a magnetometer~\cite{Mizu18}.

The significance of this work is two-fold; first, it would open up the possibility to an area of nonlinear magnetoelectric phenomena involving the nonlinear mixing between electric and magnetic fields using an atomic system. Secondly, it can be used for precision rf magnetometry by utilizing the simultaneous effect of forward four-wave mixing present in the system. If the expected sensitivity in both magnitude and the frequency resolution is reached, the system can be an ideal candidate for numerous applications such as sensing biological magnetic fields~\cite{Jens18}, detection of signals in magnetic resonance imaging (MRI) and nuclear magnetic resonance (NMR)~\cite{Xu06}, investigation of geomagnetic fields~\cite{Dang10}, measurement of the magnetic fields in space as well as for the search of dark matter~\cite{Chri16, Alex18, Chu19} and in general for the application of sensitive magnetometry in different challenging envirnments~\cite{Kai20}.

{}

\newpage
 
{\raggedright \textbf{\large Methods}}\\
{\raggedright \textbf{Theoretical Model}}:
To theoretically model the system, we consider the four-level scheme as shown in Fig.~\ref{b1} (a) and (b). The Hamiltonian of the system in the rotating frame under electric and magnetic dipole approximation is given by the expression, 
\[
{H}=-\hslash
\begin{pmatrix}
\Delta_{rf}& \Omega_{rf}& 0& \Omega_p^*\\
\Omega_{rf}^*& 0& \Omega_{rf}&   0\\
0&  \Omega_{rf}^*&  -\Delta_{rf} & 0\\
\Omega_p&   0& 0& (\Delta_p+\Delta_{rf})
\end{pmatrix}.
\]
Here, $\Omega_p (=\frac{\mu_e E_p}{\hbar})$ and $\Omega_{rf} (=\frac{\mu_m B_{rf}}{\hbar})$ are the Rabi frequencies of the input pump field and the rf field respectively. The time evolution of the system is described by the optical Bloch equation as,
\begin{equation}
\frac{d\rho}{dt}=-\frac{i}{\hbar}[H,\rho]+\mathcal{L}_{D},
\end{equation} 
with 
\[
\mathcal{L}_{D}=
\begin{pmatrix}
\Gamma\rho_{44}&
-\gamma\rho_{12}&
-\gamma\rho_{13}&
-\Gamma/2\rho_{14}\\

-\gamma\rho_{21}&
\Gamma\rho_{44} &
-\gamma\rho_{23} &
-\Gamma/2\rho_{24}\\

-\gamma\rho_{31}& 
-\gamma\rho_{32}&
\Gamma\rho_{44} &
-\Gamma/2\rho_{34}\\

-\Gamma/2\rho_{41}& 
-\Gamma/2\rho_{42}&
-\Gamma/2\rho_{43}&
-3\Gamma\rho_{44}
\end{pmatrix}.
\]

Here, $\mathcal{L}_{D}$ describes the decay and dephasing mechanism present in the system. The equations are solved for steady-state condition to evaluate the analytical expressions for $\rho_{42}$ and $\rho_{43}$, which give information about the generated field amplitudes via three and four-wave mixing processes respectively. Here, we have neglected the population in the excited state i.e. $\rho_{44}=0$ as $\Delta_p \gg \Omega_p, \Gamma, \gamma, \Delta_{rf}$. If $\Omega_p \gg \Omega_{rf}$, then the optical pumping process results in $\rho_{11}=0$ and $\rho_{22}=\rho_{33}=0.5$. Using these approximations, the susceptibilities corresponding to $\Omega_{g1}$ and $\Omega_{g2}$ are given by,

\begin{equation}
\chi^{eem}_{(g1)}= \frac{N\mu_{42}^2}{\epsilon_0\hbar\Omega_{g1}}\frac{\Omega_p \Omega_{rf}}{2\Delta_1\Delta_2} \left(\frac{c_2\Delta_2+c_0\Omega_{rf}^2}{c_1c_2-c_0^2\Omega_{rf}^2}\right), \hspace{0.5cm}
\end{equation}
\begin{equation}
\chi^{eemm}_{(g2)}= -\frac{N\mu_{43}^2}{\epsilon_0\hbar\Omega_{g2}}\frac{\Omega_p \Omega_{rf}^2}{2\Delta_1\Delta_2} \left(\frac{c_1+c_0\Delta_2}{c_1c_2-c_0^2\Omega_{rf}^2}\right).
\end{equation}

Here, $\Delta_1=\Delta_{rf}+i\gamma$, $\Delta_2=2\Delta_{rf}+i\gamma$, $c_0=1+\frac{\Omega_p^2}{2\Delta_1\Delta_2}$, $c_1=\Delta_p-\frac{\Omega_p^2}{\Delta_1}$ and $c_2=\Delta_p-\frac{\Omega_p^2}{\Delta_2}$. $N$ is the number density of the atoms whereas $\mu_{42}$ and $\mu_{43}$ are the electric dipolemoment of the corresponding atomic transitions. 

\vspace{0.5cm}
{\raggedright \textbf{Experimental Methods}}: We present the experimental setup for the observation of the mixing process in Fig.~\ref{b1}(c). The pump beam, derived from an external cavity diode laser (tuned to 780 nm) is split into two parts, where one beam is directed to a rubidium vapor cell of length 5 cm and the other part is shifted using AOMs by 40 MHz to be used as a local oscillator (LO). We use a quarter-wave (half-wave) plate to make the input pump beam circularly (linearly) polarized before the cell and another quarter-wave (half-wave) - plate to convert it back to linear (orthogonal linear) polarization after the cell. As the fields are generated along the same direction as the input pump beam, all the output beams are combined with the LO reference beam using a polarising beam splitter. This output light is then passed through a polariser and collected in a photodetector to be analyzed in a spectrum analyzer. The optical generation is optimiized when the laser is red-detuned by approximately 300 MHz to $^{87}$Rb $F = 1\rightarrow F=0$ transition.

We use Helmohtz coil-arrangement along x-, y- and z-directions around the vapor cell for applying the magnetic fields. For the experiment with input circularly polarised pump beam propagating along the z-direction, a static magnetic field is applied in the z-coil to define the quantization axis and to split the Zeeman sublevels of the ground state. In this way, the pump polarisation is always perpendicular to the dc magnetic field and hence can be made $\sigma^{+}$ or $\sigma^{-}$-polarised using the input quarter-wave plate. We apply the rf magnetic field in the y-coil and its frequency matching the Zeeman splitting. On the other hand, applying the static field along the x-coil or the y-coil leads to the realisation of any arbitrary linear polarisation for the pump beam of the form, $c_{1}|\pi\rangle+c_{2}|\sigma^{+}\rangle+c_{3}|\sigma^{-}\rangle$, where $c_{1}$, $c_{2}$ and $c_{3}$ are controlled using the input half-wave plate.  The rf field, in this case, is applied in the z-coil. The vapor cell is enclosed with three layers of $\mu-$metal sheets to shield the effect of stray magnetic fields. Furthermore, the cell is heated up to $80^0$ C, which corresponds to an atomic vapor density of $1.8 \times 10^{12}$ cm$^{-3}$. The input pump power at the entrance of the cell is 10 $\mu$W while the pump beam waist is ~$1.5$ mm.
\\

\vspace{0,5cm}

{\raggedright \textbf{Acknowledgments}}\\
The authors gratefully acknowledge the financial support by National Institute of Science Education and Research Bhubaneswar, and Tata Insititute of Fundamental Research, Department of Atomic Energy, Government of India.

 \vspace{0.5cm}
{\raggedright \textbf{Author contributions}}
SSS and AKM contributed to project planning, SSS, SRM and AKM contributed to experimental work and all the authors contributed to data analysis and preparation of the manuscript.

\end{document}